\documentclass[aps,prd,onecolumn,groupedaddress,showpacs,nofootinbib,amssymb]{revtex4}
\usepackage[dvips]{graphicx}
\usepackage{amssymb}
\usepackage{amsmath}
\usepackage{graphicx,,color}
\usepackage{amsfonts}
\usepackage{bm}
\usepackage{cancel}
\usepackage{comment}

\newcommand\be{\begin{equation}}
\newcommand\ee{\end{equation}}

\allowdisplaybreaks[4]

\begin{document}

\tolerance=5000

\title{GW170817 Viable Einstein-Gauss-Bonnet Inflation Compatible with the Atacama Cosmology Telescope Data}
\author{S.D. Odintsov$^{1,2}$}\email{odintsov@ice.csic.es}
\author{V.K. Oikonomou,$^{3,4,5}$}\email{voikonomou@gapps.auth.gr;v.k.oikonomou1979@gmail.com}
\affiliation{$^{1)}$ Institute of Space Sciences (ICE, CSIC) C. Can Magrans s/n, 08193 Barcelona, Spain \\
$^{2)}$ ICREA, Passeig Luis Companys, 23, 08010 Barcelona, Spain\\
$^{3)}$Department of Physics, Aristotle University of
Thessaloniki, Thessaloniki 54124, Greece\\
$^{4)}$L.N. Gumilyov Eurasian National University - Astana,
010008, Kazakhstan\\
$^{5)}$ Laboratory for Theoretical Cosmology, International Center
of Gravity and Cosmos, Tomsk State University of Control Systems
and Radioelectronics  (TUSUR), 634050 Tomsk, Russia}

\tolerance=5000

\begin{abstract}
In this work we investigate several Einstein-Gauss-Bonnet models
that are compatible with the GW170817 event, the Atacama Cosmology
Telescope data and the BICEP/Keck updated Planck constraints on
the tensor-to-scalar ratio. We consider two distinct classes of
Einstein-Gauss-Bonnet theories, which are equally successful for
GW170817-compatible model building and we examine their viability
against the  Atacama Cosmology Telescope data and the updated
Planck constraints on the tensor-to-scalar ratio. The two models
are distinct since the first class relates directly the
non-minimal Gauss-Bonnet scalar coupling function with the scalar
potential and yields $c_T^2=1$ while in the second class, the
non-minimal Gauss-Bonnet scalar coupling function and the scalar
potential are freely but conveniently chosen and the class of
models respects the constraint $\left| c_T^2 - 1 \right| < 6
\times 10^{-15}$. We provide several examples of models belonging
to both the two classes of GW170817-compatible
Einstein-Gauss-Bonnet theories and we demonstrate that
Einstein-Gauss-Bonnet theories provide a promising theoretical
framework for inflationary dynamics.
\end{abstract}


\maketitle

\section{Introduction}

Inflation \cite{inflation1,inflation2,inflation3,inflation4}, the
prominent theory for the description of the post-Planck classical
Universe, will be scrutinized in the next decade by ground based
cosmic microwave background (CMB) experiments
\cite{SimonsObservatory:2019qwx} and satellite gravitational wave
experiments
\cite{Hild:2010id,Baker:2019nia,Smith:2019wny,Crowder:2005nr,Smith:2016jqs,Seto:2001qf,Kawamura:2020pcg,Bull:2018lat,LISACosmologyWorkingGroup:2022jok}.
These experiments will either detect the tensor modes of the CMB,
the so-called B-modes, or the stochastic gravitational wave
background generated from inflation. The B-mode detection will
enjoy an elevated significance since it will be a smoking gun for
inflation detection. The stochastic gravitational wave background
will be a less direct detection for inflation, since other sources
may generate such a stochastic background like supermassive black
holes mergers or phase transitions and so on. The NANOGrav
detection of the stochastic gravitational wave background in 2023
\cite{NANOGrav:2023gor} does not seem to be explained by inflation
solely  \cite{Vagnozzi:2023lwo}. Recently, the reported Atacama
Cosmology Telescope data \cite{ACT:2025fju,ACT:2025tim} combined
with the DESI data \cite{DESI:2024uvr} indicated that the scalar
spectral index of the primordial scalar perturbations is
constrained as follows,
\begin{equation}\label{act}
n_{\mathcal{S}}=0.9743 \pm
0.0034,\,\,\,\frac{\mathrm{d}n_{\mathcal{S}}}{\mathrm{d}\ln
k}=0.0062 \pm 0.0052
\end{equation}
which deviates from the Planck data \cite{Planck:2018jri} at least
for 2$\sigma$. This result is strange certainly, since it disturbs
the calm waters of inflationary dynamics established in the last
decade. Since the ACT data were released, several works considered
these updates on the scalar spectral index
\cite{Kallosh:2025rni,Gao:2025onc,Liu:2025qca,Yogesh:2025wak,Yi:2025dms,Peng:2025bws,Yin:2025rrs,Byrnes:2025kit,Wolf:2025ecy,Aoki:2025wld,Gao:2025viy}.
Without judging the data, in this work we shall investigate
whether this result can be described viably in the context of
Einstein-Gauss-Bonnet (EGB) theories
\cite{Hwang:2005hb,Nojiri:2006je,Cognola:2006sp,Nojiri:2005vv,Nojiri:2005jg,Satoh:2007gn,Bamba:2014zoa,Yi:2018gse,Guo:2009uk,Guo:2010jr,Jiang:2013gza,vandeBruck:2017voa,Pozdeeva:2020apf,Vernov:2021hxo,Pozdeeva:2021iwc,Fomin:2020hfh,DeLaurentis:2015fea,Chervon:2019sey,Nozari:2017rta,Odintsov:2018zhw,Kawai:1998ab,Yi:2018dhl,vandeBruck:2016xvt,Maeda:2011zn,Ai:2020peo,Easther:1996yd,Codello:2015mba,Oikonomou:2021kql,Oikonomou:2022xoq,Odintsov:2020sqy,Oikonomou:2024etl,Fier:2025huc}.
We shall also confront these theories with the latest constraint
of the Planck data on the tensor-to-scalar ratio
\cite{BICEP:2021xfz}, which indicate that,
\begin{equation}\label{planck}
r<0.036
\end{equation}
at $95\%$ confidence. The EGB theories are plagued with a
non-trivial speed of the primordial tensor perturbations, which is
incompatible with the GW170817 observation
\cite{TheLIGOScientific:2017qsa,Monitor:2017mdv,GBM:2017lvd,LIGOScientific:2019vic},
see Refs.
\cite{Ezquiaga:2017ekz,Baker:2017hug,Creminelli:2017sry,Sakstein:2017xjx,Boran:2017rdn}.
However, the actual constraint of the GW170817 event on the
gravitational wave speed is $\left| c_T^2 - 1 \right| < 6 \times
10^{-15}$, thus there is room for a viable description of EGB
inflation, still compatible with the GW170817 event. In a series
of articles, a remedy for the EGB theories was formulated
\cite{Oikonomou:2021kql,Oikonomou:2022xoq,Odintsov:2020sqy,Nojiri:2023mbo},
which imposed the constraint $c_T^2=1$ on the primordial tensor
perturbations, resulting in a differential equation that relates
the non-minimal scalar Gauss-Bonnet function with the scalar
potential. In addition, in Ref. \cite{Oikonomou:2024etl}, the EGB
inflationary theory was developed which is compatible with the
GW170817 event constraint $\left| c_T^2 - 1 \right| < 6 \times
10^{-15}$, in which case the non-minimal scalar Gauss-Bonnet
function and the scalar potential are free to choose. In this work
we aim to study these two distinct formulations of
GW170817-compatible EGB theories in view of both the ACT data
(\ref{act}) and the updated Planck constraint (\ref{planck}). As
we demonstrate in terms of simple models, both frameworks are
perfectly compatible with both the ACT data (\ref{act}) and the
updated Planck constraint (\ref{planck}). Thus the EGB theories,
which are string theory motivated theories, provide a solid and
consistent theoretical framework that can generate a viable
inflationary phenomenology.

\section{EGB Inflation Compatible with the Constraint $\left| c_T^2 - 1 \right| < 6 \times
10^{-15}$}

\begin{figure}
\centering
\includegraphics[width=30pc]{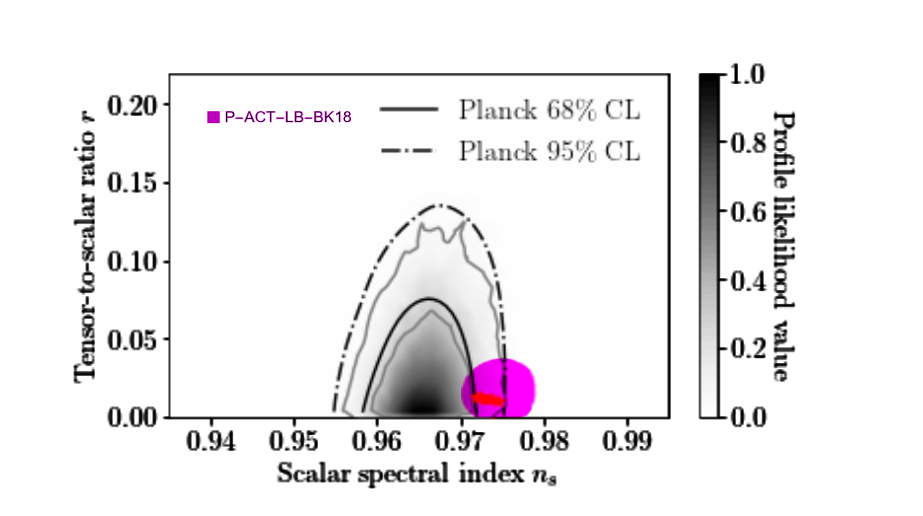}
\caption{Marginalized curves of the Planck 2018 data and RGI
potential inflation  (\ref{viablepotentials3}) confronted with the
ACT, the Planck 2018 data and the updated Planck constraint on the
tensor-to-scalar ratio.}\label{plot1}
\end{figure}

\begin{figure}
\centering
\includegraphics[width=25pc]{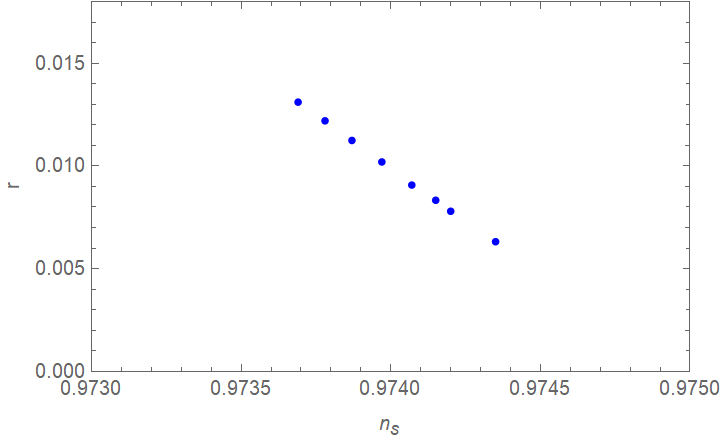}
\caption{RGI potential inflation  (\ref{viablepotentials3}): List
plot of the values for the tensor-to-scalar ratio and the spectral
index of the scalar perturbations $(n_{\mathcal{S}},r)=((0.97435,
0.006319), (0.97420, 0.0078), (0.97415, 0.008333), (0.97407,
0.009075), (0.97397, 0.01020),\\ (0.97387,0.011244), (0.973780,
0.01220), (0.97369, 0.013108))$ obtained by varying $d=[1,4]$.
}\label{refplot2}
\end{figure}

\begin{figure}
\centering
\includegraphics[width=30pc]{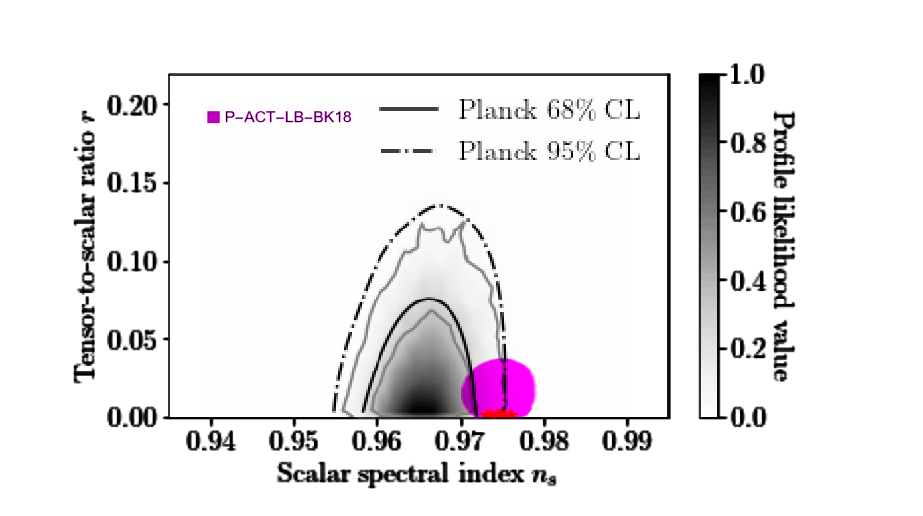}
\caption{Marginalized curves of the Planck 2018 data and IPL
potential inflation  (\ref{potential2}) confronted with the ACT,
the Planck 2018 data and the updated Planck constraint on the
tensor-to-scalar ratio.}\label{plot2}
\end{figure}

\begin{figure}
\centering
\includegraphics[width=25pc]{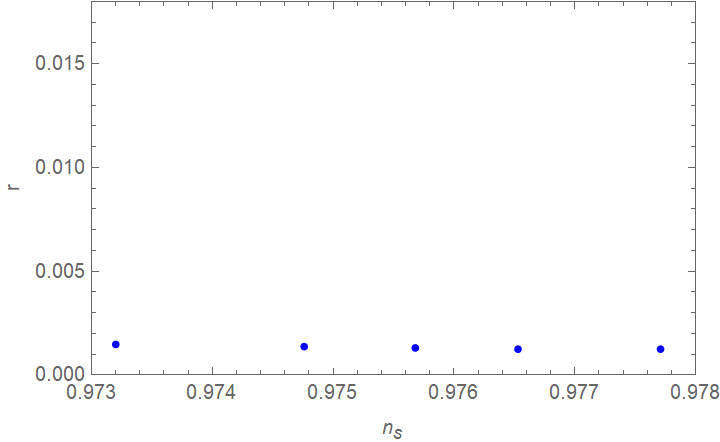}
\caption{IPL potential inflation  (\ref{potential2}): List plot of
the values for the tensor-to-scalar ratio and the spectral index
of the scalar perturbations $(n_{\mathcal{S}},r)=((0.9732,
0.00147), (0.97476, 0.001367), (0.97568, 0.001301), (0.97653,
0.001241), (0.97771, 0.001241))$ obtained by varying $N=[50,60]$.
}\label{refplot3}
\end{figure}

As we mentioned in the introduction, there are two classes of EGB
models which can provide a viable inflationary era compatible with
the GW170817 event, and these were developed in Refs.
\cite{Oikonomou:2021kql,Oikonomou:2024etl}. In this section we
shall investigate the first class of models developed in
\cite{Oikonomou:2024etl} and show that it can provide results that
are compatible with both the ACT data (\ref{act}) and the updated
Planck data on the tensor-to-scalar ratio (\ref{planck}). We shall
review the necessary formalism and present some of the viable
models. We start off analysis with the gravitational action of EGB
theories,
\begin{equation}
\label{action} \centering
S=\int{d^4x\sqrt{-g}\left(\frac{R}{2\kappa^2}-\frac{1}{2}\partial_{\mu}\phi\partial^{\mu}\phi-V(\phi)-\frac{1}{2}\xi(\phi)\mathcal{G}\right)}\,
,
\end{equation}
with $R$ denoting the Ricci scalar, $\kappa=\frac{1}{M_p}$ with
$M_p$ being the reduced Planck mass, and $\mathcal{G}$ denotes the
Gauss-Bonnet invariant in four dimensions,
$\mathcal{G}=R^2-4R_{\alpha\beta}R^{\alpha\beta}+R_{\alpha\beta\gamma\delta}R^{\alpha\beta\gamma\delta}$.
For a flat Friedmann-Robertson-Walker (FRW) metric,
\begin{equation}
\label{metric} \centering
ds^2=-dt^2+a(t)^2\sum_{i=1}^{3}{(dx^{i})^2}\, .
\end{equation}
by varying the action with respect to the scalar field and the
metric, we obtain the field equations,
\begin{equation}
\label{motion1} \centering
\frac{3H^2}{\kappa^2}=\frac{1}{2}\dot\phi^2+V+12 \dot\xi H^3\, ,
\end{equation}
\begin{equation}
\label{motion2} \centering \frac{2\dot
H}{\kappa^2}=-\dot\phi^2+4\ddot\xi H^2+8\dot\xi H\dot H-4\dot\xi
H^3\, ,
\end{equation}
\begin{equation}
\label{motion3} \centering \ddot\phi+3H\dot\phi+V'+12 \xi'H^2(\dot
H+H^2)=0\, .
\end{equation}
We also assume a slow-roll era, realized by the following
constraints,
\begin{equation}\label{slowrollhubble}
\dot{H}\ll H^2,\,\,\ \frac{\dot\phi^2}{2} \ll V,\,\,\,\ddot\phi\ll
3 H\dot\phi\, .
\end{equation}
The speed of the tensor perturbations of a FRW spacetime is,
 \cite{Hwang:2005hb},
\begin{equation}
\label{GW} \centering c_T^2=1-\frac{Q_f}{2Q_t}\, ,
\end{equation}
with $Q_f$, $F$ and $Q_b$ being equal to $Q_f=8
(\ddot\xi-H\dot\xi)$, $Q_t=F+\frac{Q_b}{2}$,
$F=\frac{1}{\kappa^2}$ and also $Q_b=-8 \dot\xi H$. The constraint
of the GW170817 event on the speed of gravitational waves is,
\begin{align}
\label{GWp9} \left| c_T^2 - 1 \right| < 6 \times 10^{-15}\, .
\end{align}
and in order to achieve this, we assume that,
\begin{equation}\label{actualgw170817constraints}
\kappa^2\dot{\xi}H\ll 1,\,\,\,\kappa^2\ddot{\xi}\ll 1\, .
\end{equation}
If the constraints above are satisfied during the inflationary
regime, then the gravitational wave speed can be small enough in
order the constraint (\ref{GWp9}) is satisfied. In order to
formalize analytical inflation in the context of EGB theories, we
further assume that $\kappa^2\dot{\xi}H^3\ll \kappa^2\dot{\phi}^2$
and $\kappa^2\ddot{\xi}H^2\ll \kappa^2\dot{\phi}^2$, that is,
\begin{equation}\label{additionalconstraints}
\kappa^2\dot{\xi}H^3\ll
\kappa^2\dot{\phi}^2,\,\,\,\kappa^2\ddot{\xi}H^2\ll
\kappa^2\dot{\phi}^2\, .
\end{equation}
In view of the above considerations, the Friedmann equation
becomes,
\begin{equation}
\label{motion5} \centering H^2\simeq\frac{\kappa^2V}{3}\, ,
\end{equation}
and the Raychaudhuri equation becomes,
\begin{equation}
\label{motion6} \centering \dot H\simeq-\frac{1}{2}\kappa^2
\dot\phi^2\, ,
\end{equation}
while the modified Klein-Gordon equation yields,
\begin{equation}
\label{motion8} \centering \dot\phi\simeq
-\frac{12\xi'(\phi)H^4+V'}{3H}\, .
\end{equation}
The slow-roll indices for EGB inflation are \cite{Hwang:2005hb},
\begin{align}
\centering \label{indices} \epsilon_1&=-\frac{\dot
H}{H^2}&\epsilon_2&=\frac{\ddot\phi}{H\dot\phi}&\epsilon_3&=0&\epsilon_4&=\frac{\dot
E}{2HE}&\epsilon_5&=\frac{Q_a}{2HQ_t}&\epsilon_6&=\frac{\dot
Q_t}{2HQ_t}\, ,
\end{align}
with $Q_a=-4\dot{\xi} H^2 $, $Q_b=-8\dot{\xi} H$,
$E=\frac{1}{(\kappa\dot\phi)^2}\left(
\dot\phi^2+\frac{3Q_a^2}{2Q_t}+Q_c\right)$, $Q_c=0$, $Q_d=0$,
$Q_e=-16 \dot{\xi} \dot{H}$, $Q_f=8\left(\ddot{\xi}-\dot{\xi}H
\right)$ and in addition $Q_t=\frac{1}{\kappa^2}+\frac{Q_b}{2}$.
Also the observables of inflation, that is, the scalar spectral
index of the primordial curvature perturbations, and the
tensor-to-scalar ratio, these are,
\begin{equation}\label{spectralindex}
n_{\mathcal{S}}=1+\frac{2 (-2
\epsilon_1-\epsilon_2-\epsilon_4)}{1-\epsilon_1}\, ,
\end{equation}
\begin{equation}\label{tensortoscalar}
r=\left |\frac{16 \left(c_A^3 \left(\epsilon_1-\frac{1}{4} \kappa
^2 \left(\frac{2
Q_c+Q_d}{H^2}-\frac{Q_e}{H}+Q_f\right)\right)\right)}{c_T^3
\left(\frac{\kappa ^2 Q_b}{2}+1\right)}\right |\, ,
\end{equation}
with $c_A$ denoting the sound speed of the scalar perturbations
which is explicitly,
\begin{equation}\label{soundspeed}
c_A=\sqrt{\frac{\frac{Q_a Q_e}{\frac{2}{\kappa ^2}+Q_b}+Q_f
\left(\frac{Q_a}{\frac{2}{\kappa
^2}+Q_b}\right)^2+Q_d}{\dot{\phi}^2+\frac{3 Q_a^2}{\frac{2}{\kappa
^2}+Q_b}+Q_c}+1}\, .
\end{equation}
The $e$-foldings number is,
\begin{equation}
\label{efolds} \centering
N=\int_{t_i}^{t_f}{Hdt}=\int_{\phi_i}^{\phi_f}\frac{H}{\dot{\phi}}d\phi\,
,
\end{equation}
with $\phi_i$ and $\phi_f$ being the scalar field values at first
horizon crossing, and at the end of the inflation. In view of Eqs.
(\ref{efolds}) and (\ref{motion8}), we have,
\begin{equation}
\label{efolds1} \centering
N=\int_{\phi_i}^{\phi_f}\frac{2H^2}{12\xi'H^4+V'}d\phi\, ,
\end{equation}
and in turn, in view of the Friedmann equation (\ref{motion5}) we
have,
\begin{equation}
\label{efoldsuncostrained} \centering
N=\int_{\phi_f}^{\phi_i}\frac{\kappa ^2 V(\phi )}{V'(\phi
)+\frac{4}{3} \kappa ^4 V(\phi )^2 \xi '(\phi )}d\phi\, .
\end{equation}
Also a viable theory must yield an amplitude of scalar
perturbations compatible with the Planck data
\cite{Planck:2018jri} which constrain it to be
$\mathcal{P}_{\zeta}(k_*)=2.196^{+0.051}_{-0.06}\times 10^{-9}$.
The amplitude of the scalar perturbations for EGB theories is
\cite{Hwang:2005hb},
\begin{equation}\label{powerspectrumscalaramplitude}
\mathcal{P}_{\zeta}(k)=\left(\frac{k \left((-2
\epsilon_1-\epsilon_2-\epsilon_4) \left(0.57\, +\log \left(\left|k
\eta \right| \right)-2+\log (2)\right)-\epsilon_1+1\right)}{(2 \pi
) \left(z c_A^{\frac{4-n_{\mathcal{S}}}{2}}\right)}\right)^2\, ,
\end{equation}
with $z=\frac{a \dot{\phi} \sqrt{\frac{E(\phi )}{\frac{1}{\kappa
^2}}}}{H (\epsilon_5+1)}$ and
$\eta=-\frac{1}{aH}\frac{1}{-\epsilon_1+1}$, all evaluated at the
first horizon crossing. Now, using the Friedmann equation
(\ref{motion5}), the Raychaudhuri equation (\ref{motion6}) and in
addition the modified Klein-Gordon equation (\ref{motion8}), we
obtain,
\begin{equation}\label{epsilon1analytic}
\epsilon_1=\frac{4}{3} \kappa ^2 \xi '(\phi ) V'(\phi
)+\frac{V'(\phi )^2}{2 \kappa ^2 V(\phi )^2}+\frac{8}{9} \kappa ^6
V(\phi )^2 \xi '(\phi )^2\, ,
\end{equation}
thus for obtaining analytic results, we must find  a convenient
form of $\xi'(\phi)$ which can significantly simplify $\epsilon_1$
in Eq. (\ref{epsilon1analytic}) and also the integral
(\ref{efoldsuncostrained}). We shall discuss several such models
below. A convenient choice is the following,
\begin{equation}\label{couplingfunctionchoices3}
\xi'(\phi)=\frac{\lambda  V'(\phi )}{V(\phi )^2}\, ,
\end{equation}
with the propagation speed of the tensor perturbations being in
this case,
\begin{equation}\label{gwspeedclass2}
c_T^2=\frac{-8 \lambda  (4 \lambda +3)^2 V(\phi ) V'(\phi )^2
V''(\phi )+10 \lambda  (4 \lambda +3)^2 V'(\phi )^4+27 V(\phi
)^4}{3 V(\phi )^2 \left(4 \lambda  (4 \lambda +3) V'(\phi )^2+9
V(\phi )^2\right)}\, ,
\end{equation}
and also the first slow-roll index becomes,
\begin{equation}\label{epsilon1formclass}
\epsilon_1=\frac{(4 \lambda +3)^2 V'(\phi )^2}{18 V(\phi )^2}\, ,
\end{equation}
and the $e$-foldings number becomes,
\begin{equation}\label{finalinitialefoldings}
N=\int_{\phi_f}^{\phi_i} \frac{V(\phi )}{\frac{4}{3} \lambda
V'(\phi )+V'(\phi )} \mathrm{d}\phi\, .
\end{equation}
Let us note that the class of models given in Eq.
(\ref{couplingfunctionchoices3}) are known have the following
feature, the coupling term grows largely when the potential
becomes smaller during inflation, implying a slow-down of
inflationary era and no reheating, see for example
\cite{vandeBruck:2016xvt}. We found many potentials that can be
compatible with both the updated Planck constraints on the
tensor-to-scalar ratio and the ACT data, for example the
following,
\begin{equation}\label{viablepotentials1}
V(\phi)=M \left(1-\frac{\delta }{\kappa  \phi }\right)\, ,
\end{equation}
\begin{equation}\label{viablepotentials2}
V(\phi)=M \left(1-\frac{\delta }{\kappa  \phi }\right)^2\, ,
\end{equation}
\begin{equation}\label{viablepotentials3}
 V(\phi)=M \sqrt{1-\frac{d}{\kappa  \phi }}\, ,
\end{equation}
\begin{equation}\label{viablepotentials3}
V(\phi)=\frac{M (\kappa  \phi )^2}{d+(\kappa  \phi )^2}\, .
\end{equation}
Note that these potentials are chosen on the basis of simplicity
and because for these potentials it is easy to obtain analytic
expressions for the observational indices. However, the two models
we shall analyze have some physical motivation and are studied in
the literature. We shall analyze some of these for demonstrational
purposes, starting with the last one given in Eq.
(\ref{viablepotentials3}) which is well known in single scalar
field literature as radion gauge inflation scalar potential (RGI)
\cite{Fairbairn:2003yx}. In this case we have in Planck units,
\begin{equation}\label{xiphi}
\xi'(\phi)=\frac{2 d \lambda }{M \phi ^3}\, ,
\end{equation}
and the first slow-roll index is,
\begin{equation}\label{slowrollindexena}
\epsilon_1=\frac{2 d^2 (4 \lambda +3)^2}{9 \phi ^2 \left(d+\phi
^2\right)^2}\, ,
\end{equation}
while the $e$-foldings number is,
\begin{equation}\label{integraln1}
N=\frac{\frac{3 d \phi ^2}{2}+\frac{3 \phi ^4}{4}}{8 d \lambda +6
d}\Big{|}_{\phi_f}^{\phi_i}\, .
\end{equation}
A viable phenomenology compatible with both the ACT data and the
updated Planck constraint on the tensor-to-scalar ratio is
obtained for $N=60$ by choosing the free parameters
$(d,\lambda,M)=(3,10^{-12},1.9474\times 10^{-10})$ in which case
we have, $n_{\mathcal{S}}=0.973871$, $r= 0.0112445$ and the
gravitational wave speed is $|c_T^2-1|=9.12899\times 10^{-16}$
while the amplitude of the scalar perturbations is
$\mathcal{P}_{\zeta}(k_*)=2.196\times 10^{-9}$. In Fig.
\ref{plot1} we present the model's predictions versus the ACT and
Planck 2018 likelihood curves, taking also into account the
updated Planck constraints on the tensor-to-scalar ratio, for
various values of the free parameters centered in
$(d,\lambda,M)=(3,10^{-13},2.583\times 10^{-10})$.
Also all the constraints (\ref{additionalconstraints}) are
satisfied for this model. The RGI potential inflation
(\ref{viablepotentials3}) is also quite flexible
phenomenologically, and in order to further highlight the
viability of this model, let us vary the parameter $d$ in the
range $\nu=[1,4]$ and keeping the rest of the parameters the same
as in previous. For $d$ in the range $\nu=[1,4]$ we get the
following sets of values for the tensor-to-scalar ratio and the
spectral index of the scalar perturbations
$(n_{\mathcal{S}},r)=((0.97435, 0.006319), (0.97420, 0.0078),
(0.97415, 0.008333), (0.97407, 0.009075), (0.97397, 0.01020),
(0.97387,0.011244),\\ (0.973780, 0.01220), (0.97369, 0.013108))$,
which we also plot in Fig. \ref{refplot2}. As we can see, the RGI
potential inflation model is quite well fitted in the ACT data.
We shall consider another viable potential for the list above, and
specifically the following,
\begin{equation}\label{potential2}
V(\phi)=M \left(1-\frac{\delta }{\kappa  \phi }\right)^2\, ,
\end{equation}
which we shall call inverse power law (IPL). For this potential we
have in Planck units,
\begin{equation}\label{xiphi2}
\xi'(\phi)=-\frac{2 \delta  \lambda  \phi }{M (\delta -\phi )^3}\,
,
\end{equation}
and also,
\begin{equation}\label{slowrollindexena2}
\epsilon_1=\frac{2 \delta ^2 (4 \lambda +3)^2}{9 \phi ^2 (\delta
-\phi )^2}\, ,
\end{equation}
while the $e$-foldings number is,
\begin{equation}\label{integraln12}
N=-\frac{3 \left(\frac{\delta  \phi ^2}{2}-\frac{\phi
^3}{3}\right)}{2 \delta  (4 \lambda
+3)}\Big{|}_{\phi_f}^{\phi_i}\, .
\end{equation}
In this case too, a viable phenomenology both compatible with the
ACT data and the updated Planck constraint on the tensor-to-scalar
ratio can be obtained for $N=55$ by choosing the free parameters
$(\delta,\lambda,M)=(0.01,10^{-13},2.1068\times 10^{-11})$ in
which case we have, $n_{\mathcal{S}}=0.975688$, $r= 0.00130174$
and the gravitational wave speed is $|c_T^2-1|=1.05845\times
10^{-17}$ while the amplitude of the scalar perturbations is
$\mathcal{P}_{\zeta}(k_*)=2.19599\times 10^{-9}$. In Fig.
\ref{plot2} we present this model's predictions versus the ACT and
Planck 2018 likelihood curves, taking again into account the
updated Planck constraints on the tensor-to-scalar ratio, for
various values of the free parameters centered in
$(\delta,\lambda,M)=(0.01,10^{-13},2.1068\times 10^{-11})$.
In this case too, all the constraints
(\ref{additionalconstraints}) are satisfied. This model is quite
flexible phenomenologically too, and in order further demonstrate
the viability of the model, we vary the $e$-foldings number $N$ in
the range $N=[50,60]$ and keeping the rest of the parameters the
same. Thus we get the following sets of values for the
tensor-to-scalar ratio and the spectral index of the scalar
perturbations $(n_{\mathcal{S}},r)=((0.9732, 0.00147), (0.97476,
0.001367), (0.97568, 0.001301), (0.97653, 0.001241), (0.97771,
0.001241))$, which we also plot in Fig. \ref{refplot3}. As it can
be seen in this case too, the model is quite well fitted in the
ACT data.

\section{Constrained EGB Inflation with $c_T^2=1$}

\begin{figure}
\centering
\includegraphics[width=30pc]{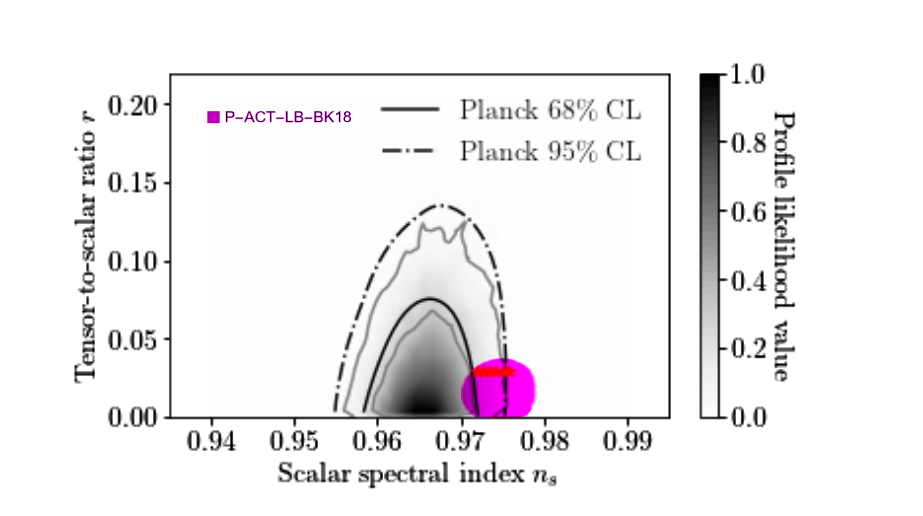}
\caption{Marginalized curves of the Planck 2018 data and the
power-law model (\ref{modelA}) confronted with the ACT, the Planck
2018 data and the updated Planck constraint on the
tensor-to-scalar ratio.}\label{plotconstrainedplanck}
\end{figure}

\begin{figure}
\centering
\includegraphics[width=25pc]{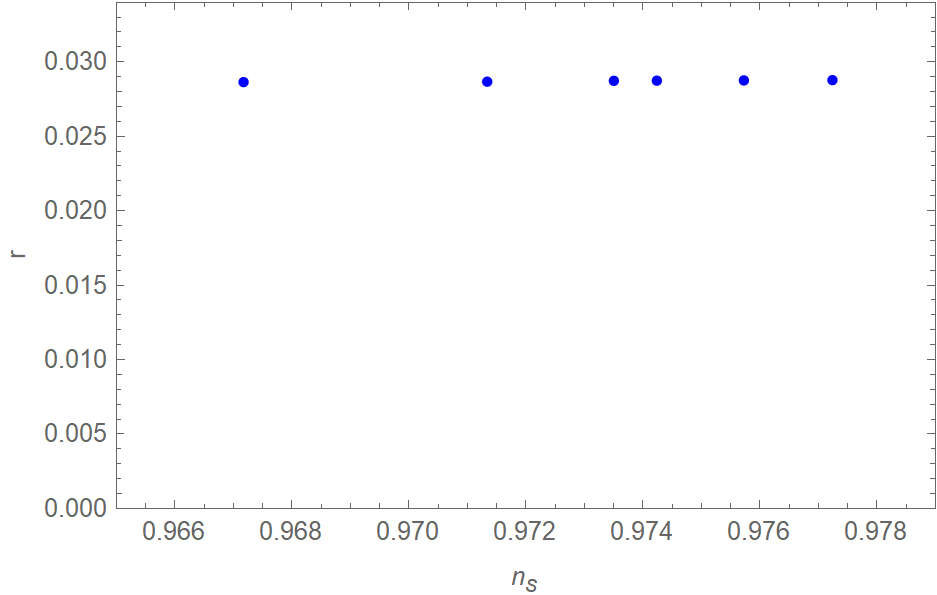}
\caption{List plot of the values for the tensor-to-scalar ratio
and the spectral index of the scalar perturbations
$(n_{\mathcal{S}},r)=((0.96717,0.02864), (0.97133,0.028671),
(0.9735,0.02872), (0.9742,0.02873),\\ (0.9757,0.02875),(0.9772,
0.02877))$ obtained by varying $\nu=[19.977,19.985]$ and for
($\beta, \gamma)=(2.89912\times 10^6,10.6\times 10^8)$ for $N=60$.
}\label{refplot1}
\end{figure}

In this section, we shall use another formalism for obtaining
viable EGB inflation compatible with the GW170817 data and
simultaneously with the latest ACT data. Specifically, consider
the EGB action of Eq. (\ref{action}) and the field equations of
Eqs. (\ref{motion1})-(\ref{motion3}) and the resulting
gravitational wave speed of Eq. (\ref{GW}). Instead of assuming
that the constraints of Eqs. (\ref{GWp9}) and
(\ref{actualgw170817constraints}) apply, we directly impose the
constraint $c_T^2=1$ which results to the condition $Q_f=0$ in Eq.
(\ref{GW}). In this case we consider the slow-roll approximation,
and the condition $Q_f=0$ results to the differential equation
$\ddot\xi=H\dot\xi$, which directly constrains the Gauss-Bonnet
scalar coupling function $\xi(\phi)$. When expressed in terms of
the scalar field, the differential equation becomes,
\begin{equation}
\label{constraintupdatedmodel1} \centering
\xi''\dot\phi^2+\xi'\ddot\phi=H\xi'\dot\phi\, ,
\end{equation}
with the ``prime'' indicating differentiation with respect to the
scalar field. By making the assumption,
\begin{equation}\label{firstslowrollupdatedmodel}
 \xi'\ddot\phi \ll\xi''\dot\phi^2\, ,
\end{equation}
motivated by the slow-roll conditions obeyed by the scalar field,
Eq. (\ref{constraintupdatedmodel1}) becomes,
\begin{equation}
\label{constraintupdatedmodel} \centering
\dot{\phi}\simeq\frac{H\xi'}{\xi''}\, .
\end{equation}
By combining Eqs. (\ref{motion3}) and
(\ref{constraintupdatedmodel}) we get,
\begin{equation}
\label{motion4updatedmodel} \centering
\frac{\xi'}{\xi''}\simeq-\frac{1}{3 H^2}\left(V'+12
\xi'H^4\right)\, .
\end{equation}
At this point we further reduce the field space by making the
assumption that we seek for models that satisfy the constraint,
\begin{equation}\label{mainnewassumptionupdatedmodel}
\kappa \frac{\xi '}{\xi''}\ll 1\, ,
\end{equation}
and furthermore this extra condition,
\begin{equation}\label{scalarfieldslowrollextraupdatedmodel}
12 \dot\xi H^3=12 \frac{\xi'^2H^4}{\xi''}\ll V\, ,
\end{equation}
being strongly related to the constraint
(\ref{mainnewassumptionupdatedmodel}). By combining Eqs.
(\ref{constraintupdatedmodel}) and
(\ref{scalarfieldslowrollextraupdatedmodel}), we can rewrite the
field equations as follows,
\begin{equation}
\label{motion5updatedmodel} \centering
H^2\simeq\frac{\kappa^2V}{3}\, ,
\end{equation}
\begin{equation}
\label{motion6updatedmodel} \centering \dot
H\simeq-\frac{1}{2}\kappa^2 \dot\phi^2\, ,
\end{equation}
\begin{equation}
\label{motion8updatedmodel} \centering
\dot\phi\simeq\frac{H\xi'}{\xi''}\, .
\end{equation}
By combining Eqs. (\ref{motion5updatedmodel}) and
(\ref{scalarfieldslowrollextraupdatedmodel}) we get,
\begin{equation}\label{mainconstraint2updatedmodel}
 \frac{4\kappa^4\xi'^2V}{3\xi''}\ll 1\, .
\end{equation}
In the context of this formalism, in contrast to the formalism
developed in the previous section, the non-minimal Gauss-Bonnet
coupling function $\xi(\phi)$ and the potential $V(\phi)$ are not
free to choose, but satisfy the following differential equation,
\begin{equation}
\label{maindiffeqnnewupdatedmodel} \centering
\frac{V'}{V^2}+\frac{4\kappa^4}{3}\xi'\simeq 0\, .
\end{equation}
The slow-roll indices in this case are,
\begin{equation}
\label{index1updatedmodel} \centering
\epsilon_1\simeq\frac{\kappa^2
}{2}\left(\frac{\xi'}{\xi''}\right)^2\, ,
\end{equation}
\begin{equation}
\label{index2updatedmodel} \centering
\epsilon_2\simeq1-\epsilon_1-\frac{\xi'\xi'''}{\xi''^2}\, ,
\end{equation}
\begin{equation}
\label{index3updatedmodel} \centering \epsilon_3=0\, ,
\end{equation}
\begin{equation}
\label{index4updatedmodel} \centering
\epsilon_4\simeq\frac{\xi'}{2\xi''}\frac{\mathcal{E}'}{\mathcal{E}}\,
,
\end{equation}
\begin{equation}
\label{index5updatedmodel} \centering
\epsilon_5\simeq-\frac{\epsilon_1}{\lambda}\, ,
\end{equation}
\begin{equation}
\label{index6updatedmodel} \centering \epsilon_6\simeq
\epsilon_5(1-\epsilon_1)\, ,
\end{equation}
with, $\mathcal{E}=\mathcal{E}(\phi)$ and $\lambda=\lambda(\phi)$
being,
\begin{equation}\label{functionE}
\mathcal{E}(\phi)=\frac{1}{\kappa^2}\left(
1+72\frac{\epsilon_1^2}{\lambda^2} \right),\,\, \,
\lambda(\phi)=\frac{3}{4\xi''\kappa^2 V}\, .
\end{equation}
Accordingly, the spectral index of the primordial tensor
perturbations $n_{\mathcal{T}}$ and the corresponding
tensor-to-scalar ratio $r$, take the following simplified forms,
\begin{equation}\label{tensorspectralindexfinalupdatedmodel}
n_{\mathcal{T}}\simeq -2\epsilon_1\left ( 1-\frac{1}{\lambda
}+\frac{\epsilon_1}{\lambda}\right)\, ,
\end{equation}
\begin{equation}\label{tensortoscalarratiofinalupdatedmodel}
r\simeq 16\epsilon_1\, ,
\end{equation}
and the $e$-foldings number becomes,
\begin{equation}
\label{efoldsupdatedmodel} \centering
N=\int_{t_i}^{t_f}{Hdt}=\int_{\phi_i}^{\phi_f}\frac{H}{\dot{\phi}}d\phi=\int_{\phi_i}^{\phi_f}{\frac{\xi''}{\xi'}d\phi}\,
.
\end{equation}
Now we can choose a simplified model for the Gauss-Bonnet scalar
coupling function, and investigate the viability of the model
confronted with the ACT data, by studying the parameter space.
Consider the model with,
\begin{equation}
\label{modelA} \xi(\phi)=\beta  (\kappa  \phi )^{\nu }\, ,
\end{equation}
with $\beta$ being dimensionless and $\kappa=1/M_p$. This model is
chosen on the basis of its simplicity, because it is very easy to
obtain analytically the scalar potential in this case. By using
(\ref{modelA}) and Eq. (\ref{maindiffeqnnewupdatedmodel}) we get
by solving the differential equation,
\begin{equation}
\label{potA} \centering V(\phi)=\frac{3}{4 \beta  \kappa ^{\nu +4}
\phi ^{\nu }+3 \gamma  \kappa ^4} \, ,
\end{equation}
with $\gamma$ being a dimensionless integration constant. Now we
easily obtain,
\begin{equation}
\label{index1A} \centering \epsilon_1\simeq \frac{\kappa ^2 \phi
^2}{2 (\nu -1)^2} \, ,
\end{equation}
\begin{equation}
\label{index2A} \centering \epsilon_2\simeq -\frac{\kappa ^2 \phi
^2-2 \nu +2}{2 (\nu -1)^2}\, ,
\end{equation}
\begin{equation}
\label{index3A} \centering \epsilon_3=0\, ,
\end{equation}
\begin{equation}
\label{index4A} \centering \epsilon_4\simeq \frac{\phi
\left(\kappa \, (2 \nu -4) \alpha (\phi ) \zeta (\phi )-8 \beta
\nu \zeta (\phi ) \kappa ^{\nu +5} \phi ^{\nu }\right)}{2 \kappa
(\nu -1) \phi  \alpha (\phi ) (\zeta (\phi )+1)} \, ,
\end{equation}
\begin{equation}
\label{index5A} \centering \epsilon_5\simeq -\frac{2 \beta \,
\kappa^4\, \nu (\kappa  \phi )^{\nu }}{(\nu -1)  \alpha (\phi )}
\, ,
\end{equation}
\begin{equation}
\label{index6A} \centering \epsilon_6\simeq -\frac{\beta \,
\kappa^4\, \nu (\kappa  \phi )^{\nu } \left(-\kappa ^2 \phi ^2+2
\nu ^2-4 \nu +2\right)}{(\nu -1)^3 \alpha (\phi )} \, .
\end{equation}
By solving $\epsilon_1\simeq \mathcal{O}(1)$ we obtain
$\phi_f\simeq \frac{\sqrt{2} (\nu-1)}{\kappa }$, and upon solving
Eq. (\ref{efoldsupdatedmodel}) with respect to $\phi_i$ we get
$\phi_i=\frac{\sqrt{2} (\nu-1) e^{-\frac{N}{\nu -1}}}{\kappa }$.
Thus the scalar spectral index becomes,
\begin{align}\label{spectralpowerlawmodelupdatedmodel}
& n_{\mathcal{S}}\simeq -1+\frac{2 (\nu -2)}{\nu -1}-2\,
e^{-\frac{2 N}{\nu -1}} + (\nu -1)^{2 \nu -3} \nu ^2 \times \\
\notag &  \frac{9\, \beta ^2 \,2^{\nu +6} \,\left(\beta \,
2^{\frac{\nu }{2}+3}\, (\nu -1)^{\nu +1}+3 \gamma  (\nu -2)
e^{\frac{\nu N}{\nu -1}}\right)}{\left(\beta \, 2^{\frac{\nu
}{2}+2}\, (\nu -1)^{\nu }+3 \gamma  e^{\frac{\nu  N}{\nu
-1}}\right)^3} \, ,
\end{align}
and the tensor-to-scalar ratio reads,
\begin{equation}\label{tensortoscalarfinalmodelpowerlawupdatedmodel}
r\simeq 16\, e^{-\frac{2 N}{\nu -1}}\, .
\end{equation}
Also, the tensor-spectral index becomes,
\begin{align}\label{tensorspectralindexpowerlawmodelupdatedmodel}
& n_{\mathcal{T}}\simeq \frac{12 \gamma  e^{\frac{(\nu -4) N}{\nu
-1}}}{\beta \, 2^{\frac{\nu }{2}+2} (\nu -1)^{\nu }+3 \gamma
e^{\frac{\nu  N}{\nu -1}}} \\ \notag & -\frac{\beta \,
2^{\frac{\nu }{2}+3} (\nu -1)^{\nu -1} e^{-\frac{4 N}{\nu -1}}
\left(-3 \nu +(\nu -1) \nu  e^{\frac{4 N}{\nu
-1}}+2\right)}{\beta\, 2^{\frac{\nu }{2}+2} (\nu -1)^{\nu }+3
\gamma  e^{\frac{\nu N}{\nu -1}}} \, .
\end{align}
Now let us investigate the parameter space to discover which
values yield viability with regard to the ACT data. A viable set
of such values  is ($\beta, \gamma,\nu)=(2.89912\times
10^6,10.6\times 10^8,19.98)$ for $N=60$. For this choice we get,
$n_\mathcal{S}=0.973503$, $n_{\mathcal{T}}=0.0442981$,
$r=0.0287288$ which are all compatible with the ACT constraints
(\ref{act}) and the update Planck constraints on the
tensor-to-scalar ratio (\ref{planck}). Also for these values of
the free parameters, the amplitude of the scalar perturbations is
$\mathcal{P}_{\zeta}(k)=2.19\times 10^{-9}$ which is compatible
with the Planck 2018 constraints \cite{Planck:2018jri}. By varying
the free parameters around the centered values ($\beta,
\gamma,\nu)=(2.89912\times 10^6,10.6\times 10^8,19.98)$ in Fig.
\ref{plotconstrainedplanck} we present the confrontation of the
present model with the combined Planck 2018 and ACT data. As it
can be seen, the present model is well fitted within the latest
ACT constraints.
Also it is easily checked that the present model respects the
constraints,
\begin{align}
& \frac{\dot\phi^2}{2} \ll V\, , \,\,\,\kappa \frac{\xi
'}{\xi''}\ll 1,\, ,\,\,\, \frac{4\kappa^4\xi'^2V}{3\xi''}\ll 1 \,
,
\end{align}
at first horizon crossing, till the end of inflation. The model is
quite flexible phenomenologically, so in order to further
highlight the viability of the model, we varied the parameter
$\nu$ in the range $\nu=[19.977,19.985]$ and keeping the rest of
the parameters the same as in previous, that is, ($\beta,
\gamma)=(2.89912\times 10^6,10.6\times 10^8)$ for $N=60$. For
various values of $\nu$ in the range $\nu=[19.977,19.985]$ we get
the following sets of values for the tensor-to-scalar ratio and
the spectral index of the scalar perturbations
$(n_{\mathcal{S}},r)=((0.96717,0.02864), (0.97133,0.028671),
(0.9735,0.02872), (0.9742,0.02873), (0.9757,0.02875),\\(0.9772,
0.02877))$, which we also plot in Fig. \ref{refplot1}. As it can
be seen, the model is quite well fitted in the ACT data.
Therefore, we demonstrated that even in this formalism, it is
possible to obtain viable EGB inflation compatible with both the
GW170817 event, the ACT data and the updated Planck constraints on
the tensor-to-scalar ratio.

\section*{Concluding Remarks}

The recent ACT data created significant noise regarding the status
quo of the constraints of the inflationary era. Without
criticizing the data, in this work we investigated whether these
data can be compatible with EGB inflation. EGB theories are well
motivated low-energy string corrected theories, and hence a
prominent candidate for inflation. However, these theories predict
a speed of primordial tensor perturbations which is distinct from
the light speed, thus contradicting at first instance the GW170817
event. We used two formulations of EGB theories that can be
compatible with the GW170817 event, with the first relating
directly the non-minimal scalar Gauss-Bonnet function with the
scalar potential and the other formulation leaving the last two
quantities unrelated. Using simple models, we demonstrated that
the refined EGB theories we considered, which are compatible with
the GW170817 event, can easily be compatible with both the ACT
data and the BICEP/Keck updates on the Planck constraints on the
tensor-to-scalar ratio. Our results indicate that EGB theories
provide a consistent, flexible and well-motivated theoretical
framework for describing inflation. It would be interesting that
such EGB gravities compatible with the GW170817 event and
consistent with ACT and BICEP/Keck updates on the Planck
constraints on the tensor-to-scalar ratio maybe combined with the
corresponding theories which pass latest LCDM/DESI observational
bounds \cite{Odintsov:2025kyw}.  The explicit construction of such
EGB gravities unifying realistic inflation with dynamical dark
energy behavior will be considered elsewhere.

\section*{Acknowledgments}

This work was partially supported by the program Unidad de
Excelencia Maria de Maeztu CEX2020-001058-M, Spain (S.D.O). This
research has been funded by the Committee of Science of the
Ministry of Education and Science of the Republic of Kazakhstan
(Grant No. AP26194585).

\end{document}